\documentstyle[12pt,wspro]{article}

\begin{document}
\rightline{SB/F/93-211}
\begin{center}
{\bf MASSIVE TRIADIC CHERN-SIMONS SPIN-3 THEORY\footnote{Talk given at
the VIII Latin American Symposium on Relativity and Gravitation,
SILARG, Sao Paulo, July 1993.}} \\ [7mm]
C. Aragone \\
Departamento de F\'{\i}sica, Universidad Sim\'on Bol\'{\i}var, \\
Apartado 8900, Caracas 1080A, Venezuela \\ [4mm]
and \\ [4mm]
A. Khoudeir \\
Departamento de F\'{\i}sica, Facultad de Ciencias, Universidad de los Andes, \\
M\'erida, 5101, Venezuela\\ [4.7cm]

\end{center}

\begin{abstract}
We introduce the massive gauge invariant, second order pure spin-3 theory in
three dimensions. It consists  of the addition of the second order gauge
invariant massless pure spin-3 action with the first order
topological(generalized) Chern-Simons spin-3 term corrected with lower spin
auxiliary actions which avoid lower spin ghosts propagation. This second order
intermediate action completes the catalogue of massive spin-3 actions having
topological structure. We also consider its spontaneous break down through the
addition of the inertial spin-3 nontopological Fierz-Pauli mass term. It is
shown that this non gauge invariant pure spin-3 system is the uniform
generalization of linearized massive vector Chern-Simons gravity and propagates
just two spin $3^{\pm}$ excitations having different masses.
\end{abstract}

\section{Introduction}

Recently an alternative curved topological gravitational theory in three
dimension was found and discussed in detail\cite{aak}. Although massive vector
Chern-Simons gravity(VCSG) propagates one massive spin-2 excitation, like
topological massive gravity(TMG)\cite{djt} substantial
differences between them occur. TMG is
described by a third order local Lorentz and diffeomorphism invariant action
whereas VCSG is a second order diffeomorphism invariant theory where
locaSNS-HEP-l Lorentz
invariance has been lost. In addition, TMG can not be broken down neither by
the presence of a triadic Chern-Simons term nor by a Fierz-Pauli(FP) mass term
while VCSG admits a symmetry breaking when the FP mass term is present, giving
rise to a double spin-2 system having two different masses\cite{sbsg}(similarly
to what happens to the
Maxwell-CS system when the Proca mass term is considered\cite{arar}). It is
worth recalling that the dreibein $e_{ma}$ is the most natural object to
describe spin-2, its gauge(linearized) transformation law is given by
$\delta_{\lambda}h_{pa} = \partial_{p}\lambda_{a}$. Self-dual massive
gravity\cite{arkh1} \cite{arkh2} completes the catalogue of spin-2 massive
excitations in three dimensions. It is a first order theory on flat Minkowski
space having no gauge invariance.

For spin-3 it is known both the topological third order theory\cite{dd} and the
self-dual first order action\cite{arkh3}. The former is gauge and local Lorentz
invariant while the latter does not have any local symmetry left. According to
what we have learnt from the spin-2 case, we will show that a gauge invariant
second-order intermediate action propagating a pure spin-$3^{+}$(or
spin-$3^{-})$ excitation indeed exists. It is the uniform generalization of
VCSG having the novelty of making unavoidable the presence of auxiliary lower
spin fields.

\section{Triadic Chern-Simons spin-3 action}

The action has the form
\begin{eqnarray}
I &=& \frac{1}{2}<h_{m\bar a\bar b}G_{m\bar a\bar b}> -
\frac{\mu}{2}<h_{m\bar a\bar
b}\epsilon^{mnp}\partial_{n}h_{p\bar a\bar b}> \\ \nonumber
&+& \mu <v_{q}\epsilon^{mnp}\partial_{p}h_{m\bar n\bar
q}> + \frac{\alpha}{2}\mu <v_{p}\epsilon^{pmn}\partial_{m}v_{n}>
+ \frac{\beta}{2}\mu^{2}<v^{2}_{p}>\\ \nonumber
&+& \mu <\phi(\partial .v)> +
\frac{\gamma}{2}<\phi\Box\phi> + \frac{\delta}{2}\mu^{2}<\phi^{2}>,
\end{eqnarray}
wherein $h_{m\bar a\bar b}$ is the basic spin-3 carrier written in the triadic
representation\cite{vas}\cite{arlrch}, symmetric and traceless in the barred
Lorentz-like indices. The core of the action consists of the usual massless
spin-3 second order action($\sim hG$) enlarged with a typical triadic
Chern-Simons first order action for spin-3($\sim h\epsilon\partial h$) which
will provide mass $\mu$. It is invariant under the natural abelian gauge
transformations $\delta_{\lambda}h_{p\bar a\bar b} = \partial_{p}\lambda_{\bar
a\bar b}$(similarly to massive vectorial Chern-Simons linearized gravity). The
remaining terms contain lower spin contributions stemming in the auxiliary
fields $v_{p}$, $\phi$ and a gauge invariant coupling term linking the vector
with the basic spin-3 field. As is well known\cite{ardy}, in $d\geq 4$, massive
higher spin fields require the presence of them to make sure the
non-propagation of undesirable lower spin ghosts. The basic triadic spin-3
field $h_{m\bar a\bar b}$ can be decomp
\begin{equation}
h_{m\bar a\bar b} = h_{\bar m\bar a\bar b} +
\epsilon_{pac}h_{\bar c\bar b}  + \epsilon_{pbc}h_{\bar c\bar a} +
\frac{3}{10}[\eta_{pa}h_{b} + \eta_{pb}h_{a} - \frac{2}{3}\eta_{ab}h_{p}].
\end{equation}

	The 15 independent components of $h_{m\bar a\bar b}$, are represented by the 7
components of $h_{\bar m\bar a\bar b}$ plus the 5 needed to describe $h_{\bar
b\bar c}$ plus the last 3 which determine $h_{p} = h_{m\bar m\bar p}$, the
unique non-vanishing trace of $h_{m\bar a\bar b}$.

	Independent variations of $h_{m\bar a\bar b}$, $v_{p}$, $\phi$ lead to the
field equations
\begin{equation}
E_{m\bar a\bar b} \equiv G_{m\bar a\bar b} -
\mu\epsilon^{mnp}\partial_{n}h_{p\bar a\bar b}
+ \frac{\mu}{2}[\epsilon^{pna}\partial_{n}v_{b} +
\epsilon^{pnb}\partial_{n}v_{a} -
\frac{2}{3}\eta_{ab}\epsilon^{pmn}\partial_{m}v_{n}] = 0,
\end{equation}
\begin{equation}
F^{p} = \alpha\mu\epsilon^{pmn}\partial_{m}v_{n} +
\beta\mu^{2}v_{p} + \mu\epsilon^{qmn}\partial_{q}h_{m\bar n\bar p} -
\mu\partial_{p}\phi = 0,
\end{equation}
and
\begin{equation}
G \equiv \mu (\partial .v) + \gamma\Box\phi + \delta\mu^{2}\phi =
0.
\end{equation}

	In order to see their dynamical content we perform a covariant analysis using
the harmonic gauge defined in this case by
\begin{equation}
\partial_{m}(h_{m\bar a\bar b} + h_{a\bar m\bar b} + h_{b\bar a\bar m}) -
(\partial_{a}h_{b} + \partial_{b}h_{a}) = 0.
\end{equation}

	Taking into account Eq.~2, this gauge fixing condition implies that
\begin{equation}
\partial_{m}h_{\bar m\bar a\bar b} =
\frac{1}{5}[\partial_{a}h_{b} + \partial_{b}h_{a} -
\frac{2}{3}\eta_{ab}\partial_{m}h_{m}],
\end{equation}
and the Eq~3 becomes
\begin{eqnarray}
\Box [ h_{m\bar a\bar b} + h_{a\bar m\bar b} + h_{b\bar a\bar m} -
\eta_{ma}h_{b} - \eta_{mb}h_{a}] &-&
\mu\epsilon^{mnp}\partial_{n}h_{p\bar a\bar b}\\ \nonumber
+ \frac{\mu}{2}[\epsilon^{pna}\partial_{n}v_{b} +
\epsilon^{pnb}\partial_{n}v_{a} -
\frac{2}{3}\eta_{ab}\epsilon^{pmn}\partial_{m}v_{n}] &=& 0.
\end{eqnarray}

	It is easy to check than none of the spin-2 variables have any dynamical
behaviour. They do not propagate. Now, let us go to the spin-1 sector. The
variables are $\rho_{m}h_{mn}^{T}, h_{p}^{T}$ and $v_{p}^{T}(\rho_{p} \equiv
\frac{\partial_{m}}{\rho}, \rho \equiv (\Box )^{\frac{1}{2}})$. The spin-1
dynamical behaviour is determined by the equations $E_{p} \equiv E_{m\bar m\bar
p} = 0, \epsilon^{mac}\partial_{c}E_{m\bar a\bar b} = 0$ and $F^{p} = 0$. In
order not to have any spin-1 excitation alive we must choose
\begin{equation}
\alpha = -\frac{27}{6}, \quad \beta = -\frac{16}{3},
\end{equation}
in order to make the spin-1 inverse propagator a non-vanishinhg number.

	Unfortunately this is not the last step in order to get a pure spin-3
propagation. There are still three scalar ghosts($\partial_{pn}h_{\bar p\bar
n}, \partial_{p}h_{p}, \partial_{p}v_{p}$) that might propagate. This is the
reason why we have to introduce the auxiliary scalar field $\phi$. The key
equations for the scalar sector are $\partial_{p}E_{p} = 0,
\epsilon^{mac}\partial_{cb}E_{m\bar a\bar b} = 0, \partial_{p}F^{p} = 0$ and $G
= 0$. One finds that
\begin{equation}
\gamma = 0, \quad \delta = -\frac{1}{6},
\end{equation}
entails the non-propagation of the whole scalar sector. Consequently, the
dynamics is contained in the symmetric, traceless, transverse part of the
triadic spin-3 field: $h_{\bar m\bar n\bar p}^{T}$. Its two independent
components can be split into parity sensitive parts according to
\begin{equation}
h^{T}_{\bar m\bar n\bar p} = h^{T+}_{\bar m\bar n\bar p} + h^{T-}_{\bar m\bar
n\bar p},
\end{equation}
with
\begin{equation}
h^{T\pm}_{\bar m\bar n\bar p} \equiv \frac{1}{2}h^{T}_{\bar m\bar
n\bar p} \pm \frac{1}{6}(\epsilon_{rsm}\rho_{r}h^{T}_{\bar
s\bar n\bar p} + \epsilon_{rsn}\rho_{r}h^{T}_{\bar
s\bar m\bar p} + \epsilon_{rsp}\rho_{r}h^{T}_{\bar
s\bar m\bar n}).
\end{equation}

	Analysing the field equation (3), we find that $h^{T-}_{\bar m\bar n\bar p} =
0$  for $\mu > 0$. An iterative process yield
\begin{equation}
(\Box - \frac{\mu^{2}}{9})h^{T+}_{\bar m\bar n\bar p} = 0,
\end{equation}
i.e. our action describes a single pure massive spin-$3^{+}$ excitation.

\section{Spontaneous break-down of translational gauge invariance}

	We want to analyse the possibility of breaking down the (translational)local
gauge invariance. We introduce a Fierz-Pauli spin-3 mass term $\sim m^{2}hh$
and consider the following action
\begin{eqnarray}
I &=& \frac{1}{2}<h_{m\bar a\bar b}G_{m\bar a\bar b}> +
\frac{\mu}{2}<h_{m\bar a\bar
b}\epsilon^{mnp}\partial_{n}h_{p\bar a\bar b}>\\ \nonumber
&-&
\frac{1}{6}m^{2}<\epsilon^{pmn}\epsilon^{abc}\eta_{pc}h_{m\bar
b\bar d}h_{n\bar c\bar d}> + \mu^{2}<h_{p}v_{p}>\\ \nonumber
&+&\frac{\mu}{2}\alpha<v_{p}\epsilon^{pmn}\partial_{m}v_{n}> +
\frac{\mu^{2}}{2}\beta<v^{2}_{p}>\\ \nonumber
&+& \mu <\phi\partial_{p}v_{p}> + \frac{\mu^{2}}{2}\delta
<\phi^{2}> + \frac{1}{2}\gamma <\phi\Box \phi>.
\end{eqnarray}
where we have taken for simplicity an algebraic(non-gauge invariant) coupling
term $\sim \mu^{2}h.v$ instead of the (gauge invariant) differential one we
introduced in the initial action.

	It is straightforward to prove, performing a covariant analysis  that
\begin{equation}
\alpha = -18(\frac{\mu}{m})^{4} = (\frac{\mu}{m})^{2}\beta, \quad \gamma = 0,
\quad \delta = \frac{m^{4}}{24(\mu^{4} + \mu^{2}m^{2})}
\end{equation}
induce the vanishing of all lower spin and our system only propagates two
spin-3 excitations represented by the traceless, transverse, parity sensitive
parts of $h_{m\bar a\bar b}^{T}$ defined in Eq.~11. Their evolution equations
are found to be
\begin{eqnarray}
\Box h^{T+}_{\bar m\bar a\bar b} &+& \frac{1}{3}\mu(\Box
)^{\frac{1}{2}}h^{T+}_{\bar m\bar a\bar b}  -
\frac{1}{9}m^{2}h^{T+}_{\bar m\bar a\bar b} = 0,\\ \nonumber
\Box h^{T-}_{\bar m\bar a\bar b} &-& \frac{1}{3}\mu(\Box
)^{\frac{1}{2}}h^{T-}_{\bar m\bar a\bar b}  -
\frac{1}{9}m^{2}h^{T-}_{\bar m\bar a\bar b} = 0,
\end{eqnarray}
giving rise to two different masses
\begin{equation}
m_{\pm} = \frac{\mu}{6}[(1 + 18\frac{m^{2}}{\mu^{2}})^{\frac{1}{2}} \mp 1].
\end{equation}
corresponding to the two spin-$3^{\pm}$ excitations that action (14)
propagates.

\section{Acknowledgements}

	One of the authors (A.K) would like to thank to the Consejo de Desarrollo
Cient\'{\i}fico y Human\'{\i}stico de la Universidad de los Andes(CDCHT-ULA) by
institutional support under project C-521-91.


\section{References}
\begin{thebibliography}{9}
\leftmargin 2.5em

\bibitem{aak}
C.~Aragone, P.~J.~Arias and A.~Khoudeir,
{\it preprint SB/F/92-192, hep-th/9307003\/}.

\bibitem{djt}
S.~Deser, R.~Jackiw and S.~Templeton,
{\it Ann. of Phys.\/} {\bf 140} (1982) 372,(E) {\bf 185} (1988).

\bibitem{sbsg}
C.~Aragone, P.~J.~Arias and A.~Khoudeir,
{\it SB/F-93-202\/}

\bibitem{arar}
C.~Aragone and P.~J.~Arias,
{\it Mod. Phys. Lett. \/} {\bf A5} (1990) 1651.

\bibitem{arkh1}
C.~Aragone and A.~Khoudeir,
{\it Phys.Lett.\/} {\bf B173} (1986) 141.

\bibitem{arkh2}
C.~Aragone and A.~Khoudeir, in {\it Quantum Mechanics of Fundamental Systems
1}, ed.\ C.~Teitelboim (Plenum Press, New York, 1988) p.\ 17.

\bibitem{dd}
T.~Damour and S.~Deser,
{\it Ann. Inst. Henri Poincar\'e \/} {\bf 47} (1987) 277.

\bibitem{arkh3}
C.~Aragone and A.~Khoudeir,
{\it Revista Mexicana de F\'{\i}sica \/} {\bf 6} (1993).

\bibitem{vas}
M.~A.~Vasiliev,
{\it Sov. J. Nucl. Phys. \/} {\bf 32} (1980) 439.

\bibitem{arlrch}
C.~Aragone and H.~La Roche,
{\it Nuovo Cimento \/} {\bf A72} (1982) 149.

\bibitem{ardy}
C.~Aragone, S.~Deser and Z.~Yang,
{\it Ann. of Phys. \/} {\bf 179} (1987) 76.


\end{thebibliography}
\end{document}